\documentclass[12pt]{article}

\usepackage{ graphicx, setspace, amsmath, amssymb, booktabs, verbatim, bm,threeparttable, amsthm, color}
\usepackage[round]{natbib}
\usepackage{algorithm}
\newtheorem{theorem}{Theorem}[]

\allowdisplaybreaks[3]

\newtheorem{corollary}{Corollary}

\begin{document}
\title{\vskip -30pt Geometric standardized mean difference and its application to meta-analysis}
\author{\small Jiandong Shi$^1$, Xiaochen Zhang$^2$\thanks{Corresponding author. E-mail: zhangxiaochen@bnu.edu.cn}, Lu Lin$^3$, Hiu Yee Kwan$^4$ and Tiejun Tong$^5$\thanks{Corresponding author. E-mail: tongt@hkbu.edu.hk} \\ \\
	{\small $^1$Department of Statistics, The Chinese University of Hong Kong, Hong Kong}\\
	{\small  $^2$Faculty of Arts and Sciences, Beijing Normal University, Zhuhai, China}\\
	{\small $^3$Zhongtai Securities Institute for Financial Studies, Shandong University, Jinan, China}\\
	{\small $^4$School of Chinese Medicine, Hong Kong Baptist University, Hong Kong}\\
	{\small $^5$Department of Mathematics, Hong Kong Baptist University, Hong Kong}\\
}		
\date{}
\maketitle

\begin{abstract}
	\baselineskip 16pt
	\noindent
	The standardized mean difference (SMD) is a widely used measure of effect size, 
	particularly common in psychology, clinical trials, and meta-analysis involving continuous outcomes. 
	Traditionally, under the equal variance assumption, the SMD is defined as the mean difference divided by a common standard deviation. 
	This approach is prevalent in meta-analysis but can be overly restrictive in clinical practice. 
	To accommodate unequal variances, the conventional method averages the two variances arithmetically, 
	which does not allow for an unbiased estimation of the SMD. 
	Inspired by this, we propose a geometric approach to averaging the variances, 
	resulting in a novel measure for standardizing the mean difference with unequal variances. 
	We further propose the Cohen-type and Hedges-type estimators for the new SMD, 
	and derive their statistical properties including the confidence intervals.
	Simulation results show that the Hedges-type estimator performs 
	optimally across various scenarios, demonstrating lower bias, lower mean squared error, and improved coverage probability. A real-world meta-analysis also illustrates that
	our new SMD and its estimators provide valuable insights to the existing literature 
	and can be highly recommended for practical use. 
	\\
	
	\noindent
	$Key \ words$:
	Effect size,
	Geometric averaging,
	Meta-analysis,
	Standardized mean difference,
	Unbiased estimator,
	Unequal variances
\end{abstract}

%\linespread{1.5}
\linespread{1.65}
\baselineskip 20pt
\setlength{\parskip}{0.3\baselineskip}

%%%%%%%%%%%%%%%%%%%%%%%%%
\newpage
\section{Introduction}\label{introduction}
\noindent
Effect size has emerged as a critical complement to statistical significance in quantitative research, reflecting its importance championed by leading scientific bodies. 
The American Psychological Association (\citealp{Wilkinson1999}) and the American Statistical Association (\citealp{Wasserstein2016}) have advocated for effect size reporting to enrich the interpretation of research findings. Within case-control studies involving continuous outcomes, the mean difference (MD) serves as a natural and intuitive effect size metric, particularly when outcomes share identical measurement scales (\citealp{Borenstein2009}). 
This direct comparability is exemplified by universally standardized measures such as blood pressure (millimeters of mercury, mmHg), where MD yields immediately interpretable results that readily support cross-study synthesis such as the systematic review and meta-analysis.

However, the measures of an effect may not always be consistent in practice. When outcome measures employ divergent scales across studies, raw MD values become fundamentally incomparable, rendering direct aggregation statistically untenable. 
For example, in the intelligence assessment, 
different studies utilize distinct psychometric instruments (e.g., Wechsler adult intelligence scale (WAIS), Stanford-Binet) that generate non-equivalent score metrics (\citealp{Fancher1985,Kaufman2006}), precluding meaningful MD comparison and synthesis. 
Another example is that the depression measurement employs disparate scales like the Beck Depression Inventory (BDI) and Hamilton Depression Rating Scale (HDRS) (\citealp{Noetel2024}), each with unique structural properties. 
The BDI's 21-item scale (0–3 per item) yields a maximum score of 63 (\citealp{Beck1961}), while the HDRS's variable structure (17–29 items scored 0–5) may produce maxima exceeding 80 (\citealp{Hamilton1960}). 
Critically, these instruments establish different clinical thresholds and numerical ranges, creating systematic bias when MDs are naively combined. 
In meta-analysis, studies using wider-ranged scales (e.g., HDRS) would disproportionately influence pooled estimates due to their larger potential MD magnitudes, possibly yielding clinically misleading conclusions. 
Consequently, methodological standardization becomes essential to place all effect sizes on a common scale for the subsequent synthesis.

To deal with the problem, the standardized mean difference (SMD) is introduced in the literature as 
\begin{equation}\label{smd_general}
	{\rm SMD} = \dfrac{{\rm mean\ of\ case\ group}-{\rm mean\ of\ control\ group}}{\rm standardizer},
\end{equation}
where the standardizer is to standardize the mean difference so that the resulting SMD has a uniform scale
for different studies, no matter whether the raw data are measured under different scales (\citealp{cohen1969statistical,coe2002It}).
Owing to this nature, SMD perfectly matches the scope of the systematic review and meta-analysis
when possibly different scales of data are collected from individual studies (\citealp{Cooper2019}). 
Naturally, SMD plays an indispensable role therein (\citealp{Gallardo2024}).
To further explore the standardizer in (\ref{smd_general}), let the two samples from the case and control groups, respectively,  
follow the normal distributions $N(\mu_1, \sigma_1^2)$ and $N (\mu_0, \sigma_0^2)$. 
When $\sigma_1^2=\sigma_0^2=\sigma^2$,
it is natural to take $\sigma$ as the standardizer, 
yielding the most commonly used SMD as 
\begin{equation}\label{smd_equal_variance}
	\delta=\dfrac{\mu_1-\mu_0}{\sigma}.
\end{equation}

To estimate $\delta$ which is often unknown in practice, we further let $(\bar X_1,S_1^2)$ and $(\bar X_0,S_0^2)$ be the sample means and variances of the two groups with $n_1$ and $n_0$ sample sizes,
and $S_{\text{pool}}=\sqrt{[\left(n_1-1\right)S_1^2+\left(n_0-1\right) S_0^2]/(n_1+n_0-2)}$ be the pooled sample standard deviation.
\cite{cohen1969statistical} proposed to estimate $\delta$ by
\begin{equation}\label{cohend}
	d=\dfrac{\bar{X}_1-\bar{X}_0}{S_{\text{pool}}}.
\end{equation}%\sqrt{\dfrac{\left(n_1-1\right)S_1^2+\left(n_0-1\right) S_0^2}{n_1+n_0-2}}
Cohen's $d$ is simple and practical, yet as is well known, 
this estimator is often biased especially for small sample sizes. 
To reduce the bias, \cite{hedges1981distribution} introduced a bias correction factor $J_{\nu}=(2/\nu)^{1/2}\Gamma\left(\nu/2\right)/\Gamma\left((\nu-1)/2\right)$,
where $\Gamma(\alpha)=\int_0^{\infty} x^{\alpha-1} e^{-x} d x$ is the gamma function.
Under the normality assumption and equal variance, 
the unbiased estimator of $\delta$, known as Hedges' $g$, is given as
\begin{equation}\label{Hedges}
	g=d\cdot J_{n_1+n_0-2}.
\end{equation}
For practical application, \cite{hedges1981distribution} further applied $1-3/(4\nu-1)$ to approximate $J_{\nu}$ with an acceptable accuracy. 
To conclude, with equal variance, 
the SMD can naturally be defined as $\delta$ in (\ref{smd_equal_variance}), 
and moreover it can also be readily estimated by Cohen's $d$ or Hedges' $g$.
Note that Cohen's $d$ and Hedges' $g$ are prevalent in systematic review and meta-analysis (\citealp{Higgins2008}), 
but the equal variance assumption can be overly restrictive in clinical practice (\citealp{hedges2025}).

On the contrary, when the variances are not equal, the standardizer in (\ref{smd_general})
may not be immediately available for computing the SMD. 
To deal with this problem, a few measures have been proposed in the literature
\citep{glass1976primary,glass1977integrating,glass1979meta,cohen1988statistical,huynh1989unified,kulinskaya2007confidence,keselman2008generally,cumming2013cohen,shieh2013confidence,peng2014beyond,goulet2018review,aoki2020effect,delacre2021hedges,Bird2023,Suero2023}, with most of them taking a weighted arithmetic mean of the two variances in the standardizer.
Consequently, the SMD with unequal variances can often be formulated as 
\begin{equation}\label{smd_weighted}
	\delta^*_w = \dfrac{\mu_1-\mu_0}{\sqrt{w\sigma_1^2+(1-w)\sigma_0^2}},
\end{equation}
where $w\in [0,1]$ is the weight assigned to the case group.
Popular choices of $w$ include, for example, 0, 0.5, 1 and $n_0/(n_1+n_0)$.
A main motivation for proposing (\ref{smd_weighted}) is 
to facilitate the estimation and inference by noting that the weighted sample variance $wS_1^2 + (1-w)S_0^2$
approximately follows a chi-squared distribution.
However, this approximation leads to an obvious limitation that an exactly unbiased estimator of $\delta^*_w$ may not exist for any $w\in (0,1)$. 
In addition, the possibly low coverage probability of the confidence interval (CI) is also a concern. 
On the other hand, if we take $w=0$ or $w=1$,  
it will then suffer from another limitation that the variance from one group 
will be completely ignored when computing the SMD.
For more details, see Section \ref{heteroscedasticity_review}.

In view of the limitations on the existing SMDs, 
we propose a new measure for standardizing the mean difference allowing for unequal variances. 
Rather than to take the weighted arithmetic mean in (\ref{smd_weighted}),
we consider a geometric approach to weight the two variances,
yielding an alternative definition of SMD as
\begin{equation}\label{new_smd}
	\delta_w=\dfrac{\mu_1-\mu_0}{\sqrt{(\sigma_1^2)^w(\sigma_0^2)^{1-w}}} =  \dfrac{\mu_1-\mu_0}{\sigma_1^w\sigma_0^{1-w}} ,
\end{equation}
where $w$ is the power weight assigned to the case group. 
It is noteworthy that the two SMDs in (\ref{smd_weighted}) and (\ref{new_smd}) will be the same when $w=0$ or 1,
and both of them reduce to $\delta=(\mu_1-\mu_0)/\sigma$ as defined
in (\ref{smd_equal_variance}) when the two variances are equal.
With the new definition, an unbiased estimator of the new measure can be derived by a bias correction to the plug-in estimator with the mutually independent sample statistics $\bar{X}_1-\bar{X}_0$, $S_1^{-w}$ and $S_0^{-(1-w)}$.
From this perspective,  our geometric SMD in (\ref{new_smd}) may turn out to be a better choice than the arithmetic SMD in (\ref{smd_weighted}) for the practical implementation.  
We propose both Cohen-type and Hedges-type estimators for the new SMD, 
and derive their statistical properties including the confidence intervals.
For more details, see Section \ref{main_results}.

The rest of the paper is organized as follows. 
Section \ref{heteroscedasticity_review} reviews the existing measures of SMD under unequal variances as well as presents their respective estimators. 
Section \ref{main_results} introduces the new definition of SMD, and provides the Cohen-type and Hedges-type estimators together with their CIs.
Section \ref{simulations} conducts simulation studies to verify that the Hedges-type estimator
performs optimally across various scenarios, demonstrating lower bias, lower mean squared error, and improved coverage probability. 
Section \ref{real_data} analyzes a real-world meta-analysis
to illustrate the practical usefulness of our new SMD. 
Lastly, the paper is concluded in Section \ref{discussions} and the technical results are presented in the Appendices.

%%%%%%%%%%%%%%%%%%%%%%%%%
\section{A review of SMDs allowing for unequal variances}\label{heteroscedasticity_review}
\noindent 
This section reviews the existing literature on standardizing the mean difference 
when the equal variance assumption does not hold. 
And as mentioned earlier, most existing SMDs can be written as a form of $\delta^*_w = (\mu_1-\mu_0)/\sqrt{w\sigma_1^2+(1-w)\sigma_0^2}$,
with the key difference on how to assign the weight $w\in [0,1]$. 

\subsection{SMD with $w=0$ or $1$}\label{glass}
\noindent
\cite{glass1981} recommended to take the standard deviation of the data from  either group as the standardizer in (\ref{smd_general}).
Without loss of generality, by taking $w=0$, it results in the SMD as
$$\delta_0^*=\dfrac{\mu_1-\mu_0}{\sigma_0}.$$
To further estimate $\delta_0^*$, the plug-in estimator is $d_0^*=(\bar{X}_1-\bar{X}_0)/S_0$
and the bias-corrected estimator is $g_0^* = J_{n_0-1}\cdot d_0^*$, 
where $J_{n_0-1}$ is the bias correction factor as defined in Section \ref{introduction}. 
For ease of reference, we also refer to them as the Cohen-type and Hedges-type estimators, respectively.
Moreover, \cite{Algina2006} showed that
\begin{equation*}\label{algina}
	\left(\dfrac{1}{n_0}+\dfrac{\sigma_1^2}{\sigma_0^2n_1}\right)^{-1/2} d_0^*\sim t(n_0-1,\lambda),
\end{equation*}
where $t(n_0-1,\lambda)$ is the noncentral $t$-distribution with $n_0-1$ degrees of freedom and noncentrality parameter $\lambda=(\mu_1-\mu_0)/\sqrt{\sigma_1^2/n_1+\sigma_0^2/n_0}$.
Letting also $\hat{\lambda}=(\bar{X}_1-\bar{X}_0)/[\sqrt{S_1^2/n_1+S_0^2/n_0}]$,
a $100(1-\alpha)\%$ CI of $\delta_0^*$ can be constructed as 
\begin{equation}\label{ci_algina}
	\left[\sqrt{\dfrac{1}{n_0}+\dfrac{S_1^2}{S_0^2n_1}}\cdot t_{1-\alpha/2}(n_0-1,\hat{\lambda}),\ \sqrt{\dfrac{1}{n_0}+\dfrac{S_1^2}{S_0^2n_1}}\cdot t_{\alpha/2}(n_0-1,\hat{\lambda})\right],
\end{equation}
where $t_{\alpha/2}(n_0-1,\hat{\lambda})$ represents the upper $\alpha/2$ quantile of the noncentral $t$-distribution
with $n_0-1$ degrees of freedom and noncentrality parameter $\hat{\lambda}$.

A main advantage of the SMD with $w=0$ or 1 is its simplicity for practical use, 
together with the availability of an unbiased estimator for any fixed sample sizes. 
On the other hand, this simplified SMD also has two obvious limitations. 
First, by taking the standard deviation of the data from only one group, 
the SMD cannot reflect the data variation from the other group. 
Second, the $100(1-\alpha)\%$ CI in (\ref{ci_algina}) may suffer from low coverage probability
especially for small sample sizes.

\subsection{SMD with $w\in(0,1)$}\label{smd01}
\noindent
Other than the 0 and 1 weights, 
\cite{cohen1988statistical} recommended to apply the equal weight $w=0.5$, 
yielding $\delta_{0.5}^* = (\mu_1-\mu_0)/\sqrt{(\sigma_1^2+\sigma_0^2)/2}$.
For more details on this SMD, one may refer to \cite{huynh1989unified} and \cite{keselman2008generally}.
%We denote this definition of SMD as $\delta_{0.5}$ and $g$ with $w=1/2$ as $g_{0.5}$.
In addition, \cite{aoki2020effect} considered the weight $w=n_1^{-1}/(n_1^{-1}+n_0^{-1})$, or equivalently $w=n_0/(n_1+n_0)$, 
by following a similar idea as Welch's $t$-test (\citealp{welch1938significance}).
One problem of Aoki's SMD is, however, that it will depend on the study sample sizes, 
which hence affects the subsequent data analysis including the meta-analytical results \citep{cumming2013cohen}. 

For any SMD in this category, including $\delta_{0.5}^*$ and $\delta_{n_0/(n_1+n_0)}^*$, 
the Cohen-type estimator is given as
\begin{equation}\label{dw*}
	d_w^*=\dfrac{\bar{X}_1-\bar{X}_0}{\sqrt{wS_1^2+(1-w)S_0^2}}.
\end{equation}
Nevertheless, since the weighted sample variance does not follow a chi-square distribution,
the Cohen-type estimator cannot yield a fully unbiased estimator even after the bias correction is implemented. 
As a compromise, an approximate Hedges-type estimator is known as
\begin{equation}\label{g_w*}
	g^*_w=d_w^*\cdot J_{\hat{\nu}_w},
\end{equation}
where $\hat{\nu}_w=[(wS_1^2+(1-w)S_0^2)^2]/[w^2S_1^4/(n_1-1)+(1-w)^2S_0^4/(n_0-1)]$ is the approximated degrees of freedom derived from the Welch-Satterthwaite equation (\citealp{welch1938significance,Satterthwaite1941}). 
In Section \ref{comparison_ari},
we will show that this Hedges-type estimator will always result in a non-negligible bias 
unless the sample sizes are sufficiently large. 

Letting $\nu_w=(w\sigma_1^2+(1-w)\sigma_0^2)^2/[w^2\sigma_1^4/(n_1-1)+(1-w)^2\sigma_0^4/(n_0-1)]$, 
we further show in Appendix A that, for any given $w\in (0,1)$, 
\begin{equation}\label{huynh}
	\left(\dfrac{\sigma_1^2/n_1+\sigma_0^2/n_0}{w\sigma_1^2+(1-w)\sigma_0^2}\right)^{-1/2}\cdot \dfrac{\bar{X}_1-\bar{X}_0}{\sqrt{wS_1^2+(1-w)S_0^2}}\sim t(\nu_w,\lambda)\quad {\rm approximately}.
\end{equation}
Moreover, by following the similar arguments as in Section \ref{glass}, the $100(1-\alpha)\%$ CI for $\delta^*_w$ can be constructed as
\begin{equation}\label{ci_general}
	\left[\sqrt{\dfrac{S_1^2/n_1+S_0^2/n_0}{wS_1^2+(1-w)S_0^2}}\cdot t_{1-\alpha/2}(\hat{\nu}_w, \hat{\lambda}),\ \sqrt{\dfrac{S_1^2/n_1+S_0^2/n_0}{wS_1^2+(1-w)S_0^2}}\cdot t_{\alpha/2}(\hat{\nu}_w,\hat{\lambda})\right],
\end{equation}
which, however, also suffers from low coverage probability (\citealp{bonett2008confidence}). 
Note that for the special cases of $w=0.5$ and $w=n_0/(n_1+n_0)$, the corresponding results 
had already been derived by \cite{huynh1989unified} and \cite{aoki2020effect}, and they coincide with what we derived in (\ref{huynh}) and (\ref{ci_general}).
Another way to establish the CI is by the asymptotic normality of $d_w^*$
as well as its large sample variance calculated by the properties of the noncentral $t$-distribution.
For the special case of $w=0.5$, \cite{bonett2008confidence} proposed an approximate normal CI
with the approximated variance to improve the performance.
      
%%%%%%%%%%%%%%%%
\section{New SMD and its estimation}\label{main_results}
\noindent
As is known, the existing SMD in (\ref{smd_weighted}) cannot be unbiasedly estimated for any $w\in (0,1)$, mainly because the weighted sample variance does not follow a chi-square distribution. 
To overcome the problem, 
we now take a geometric, rather than arithmetic, approach to average the two unequal variances. 
More specifically, our new definition of SMD is
\begin{equation}\label{smd_new}
	\delta_{w}= \dfrac{\mu_1-\mu_0}{\sigma_1^w\sigma_0^{1-w}},
\end{equation}
where $w\in [0,1]$ is the weight assigned to the case group, 
as already seen in Section \ref{introduction}. 
With the geometric weighting approach, our new SMD remains invariant for
any linear transformation $f(x)=ax+b$ with $a>0$
to both the case and control groups.
Note also that our new SMD will reduce to the Glass' SMD in Section \ref{glass} when $w=0$ or 1, and to the pooled SMD in (\ref{smd_equal_variance}) when $\sigma_1^2 =\sigma_0^2 = \sigma^2$. 
Moreover, $\delta_w$ can also be thought of as a weighted product of  $\delta_1=(\mu_1-\mu_0)/\sigma_1$ and $\delta_0=(\mu_1-\mu_0)/\sigma_0$, 
which are two simplified SMDs very easy to handle.

\vspace{0.4cm}
\begin{figure}[H]
	\centering
	\includegraphics[width=0.9\textwidth]{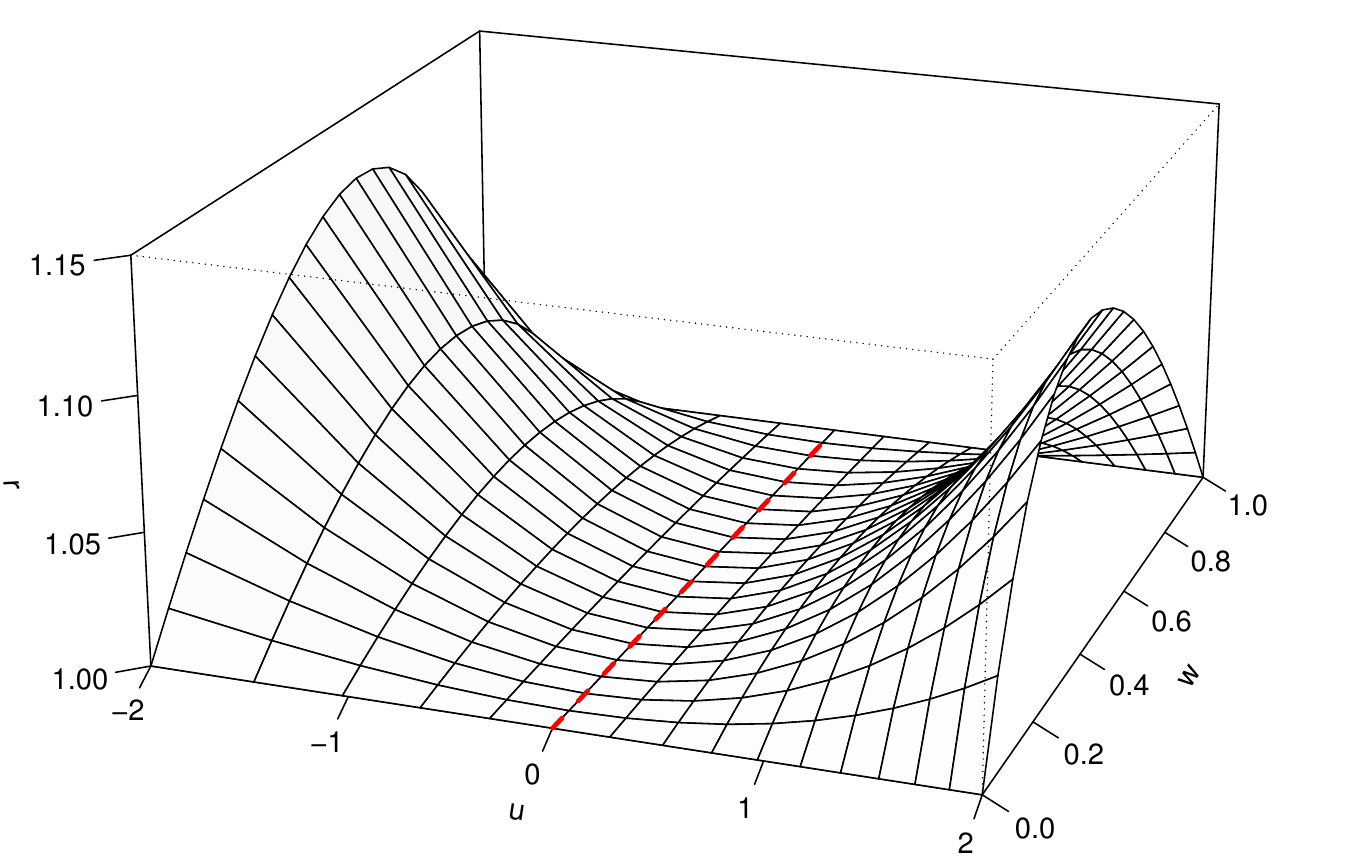}
	\setlength{\abovecaptionskip}{-10pt}%
	\setlength{\belowcaptionskip}{0pt}%
	\caption{The ratio $r=\delta_w/\delta_w^*$ varies with $u=\log_2(\sigma_1^2/\sigma_0^2)$ and $w$. The red dashed line represents the equal variance case with $u=0$, or equivalently, $\sigma_1^2=\sigma_0^2$.}
	\label{fig:ratio}
\end{figure}

For further comparison, the geometric SMD is always larger than or equal to the arithmetic SMD, i.e., $\delta_w\geq \delta_w^*$, 
by noting that the weighted geometric mean is always smaller than or equal to the weighted arithmetic mean such that $(\sigma_1^2)^w(\sigma_0^2)^{1-w}\leq w\sigma_1^2+(1-w)\sigma_0^2$.
Additionally, when $\mu_1\neq \mu_0$, we also present their ratio $r=\delta_w/\delta_w^*$ in Figure \ref{fig:ratio}
according to the degree of unbalance on the two variances, represented by $u=\log_2(\sigma_1^2/\sigma_0^2)$, together with the different choice of the weight $w\in[0,1]$.
From Figure \ref{fig:ratio}, it is evident that the ratio $r$ will never fall below 1. 
Also as expected, the two SMDs will provide the same measure when the two variances are equal and/or when the weight is 0 or 1. 
In contrast, the difference between the two SMDs will increase when the two variances become more unbalanced, in particular if $w$ is also in the middle of $[0,1]$. 
To conclude, the two SMDs are able to provide a similar measure for standardizing the mean difference when the two variances do not differ too much. 
Otherwise,  it is highly recommended to first apply the domain knowledge 
to determine which SMD should be adopted for their specific study. 
And from this perspective, our new SMD has, in fact, 
provided a useful alternative to complement the existing literature.

\subsection{Cohen-type estimation}\label{section_plug_in}
\noindent 
To estimate the newly defined SMD  in (\ref{smd_new}), we first apply the Cohen-type (plug-in) method owing to its simplicity. 
Specifically, for any given $w \in [0,1]$, our Cohen-type estimator of $\delta_w$ is given as
\begin{equation}\label{plug-in}
	d_{w}=\dfrac{\bar{X}_1-\bar{X}_0}{S_1^wS_0^{1-w}},
\end{equation}
where $\bar X_1$, $\bar X_0$, $S_1$ and $S_0$ are the sample means and standard deviations of the two groups as in Section \ref{introduction}. 
Moreover, by Theorem \ref{theorem1} in Appendix B, 
$d_w$ is asymptotically normally distributed with mean $\delta_w$ and variance 
\begin{equation}\label{var_dw1}
	{\rm Var}(d_w)\approx \dfrac{\delta_{w}^2}{2}\left(\dfrac{w^2}{n_1}+\dfrac{(1-w)^2}{n_0}\right) +\dfrac{n_1(\sigma_0^2/\sigma_1^2)^w+n_0(\sigma_1^{2}/\sigma_0^2)^{1-w}}{n_1n_0}.
\end{equation}

In the special case when $\sigma_1^2 = \sigma_0^2$, we have $\delta_w = \delta$ 
so that our new $d_w$ and Cohen's $d$ are to estimate the same quantity. 
For their numerical performance, see Section \ref{comparison_dg}.
If we further take $w=n_1/(n_1+n_0)$,
then by \cite{Hedges1985} and \cite{Borenstein2009}, 
the two estimators will be asymptotically the same by noting that
\begin{equation}\label{vard}
	{\rm Var}(d)\approx \dfrac{\delta^2}{2(n_1+n_0)}+\dfrac{n_1+n_0}{n_1n_0}.
\end{equation}
More generally, when $\sigma_1^2\neq \sigma_0^2$, 
the pooled SMD in (\ref{smd_equal_variance}) cannot be defined so that Cohen's $d$ will also be  meaningless. 
If, instead, we apply Cohen's $d$ to estimate our newly defined SMD in (\ref{smd_new}), 
then as will be shown in Section \ref{comparison_dg}, 
its performance will quickly get worse when the variances and/or the sample sizes become more unbalanced (\citealp{hedges2025}).

Apart from the asymptotic variance of $d_w$ in (\ref{var_dw1}),
when the sample sizes are small, Theorem \ref{theorem2} in Appendix C shows that the exact variance of $d_w$ has a rather complicated form and may not be easy to implement in practice. 
To overcome the problem, we further show in Appendix D that
the large sample formula in (\ref{var_dw1}) can be, in fact, adjusted to  approximate the exact variance 
by replacing $n_1$ and $n_0$ with $n_1-1$ and $n_0-1$, respectively.
In light of this, a unified $100(1-\alpha)\%$ Cohen-type CI of $\delta_{w}$ for any sample sizes can then be constructed as
$[d_{w}-z_{\alpha/2}\widehat{{\rm SE}}_{d_{w}},\ d_{w}+z_{\alpha/2}\widehat{{\rm SE}}_{d_{w}}],$
where 
\begin{equation}\label{se_dw}
	\widehat{{\rm SE}}_{d_{w}}=\sqrt{\dfrac{d_{w}^2}{2}\left(\dfrac{w^2}{n_1-1}+\dfrac{(1-w)^2}{n_0-1}\right) +\dfrac{(n_1-1)(S_0^2/S_1^2)^w+(n_0-1)(S_1^{2}/S_0^2)^{1-w}}{(n_1-1)(n_0-1)}}
\end{equation}
is the estimated standard error of $d_w$, and
$z_{\alpha/2}$ is the upper $\alpha/2$ quantile of the standard normal distribution.

To conclude, our Cohen-type estimator $d_w$ provides a very simple formula for estimating the geometric SMD. Moreover, together with its well behaved $100(1-\alpha)\%$ CI, 
our Cohen-type estimation can be highly recommended for practical use 
in both balanced and unbalanced settings expect for the extremely small sample sizes. 

\subsection{Hedges-type estimation}\label{hedges_type_w}
\noindent 
Following the previous section, to handle the small sample sizes that may yield a non-negligible bias, we further apply the Hedges-type (bias-corrected) method to propose an unbiased estimator of the geometric SMD in (\ref{smd_new}). 
Specifically, following Theorem \ref{theorem2} that
$E(d_{w})=\delta_{w}/(B_{n_1-1,w}B_{n_0-1,1-w})$, 
our Hedges-type estimator of $\delta_{w}$ is given as
\begin{equation}\label{s}
	g_w=d_w\cdot B_{n_1-1,w}B_{n_0-1,1-w},
\end{equation}
where $d_w=(\bar{X}_1-\bar{X}_0)/(S_1^wS_0^{1-w})$ is the Cohen-type estimator and $B_{\nu,w}=(2/\nu)^{w/2}$ $\Gamma(\nu/2)/\Gamma((\nu-w)/2)$ is the bias correction factor.
In addition, noting that ${\rm SE}(g_w)=B_{n_1-1,w}B_{n_0-1,1-w}\cdot {\rm SE}(d_w)$, we can further construct the $100(1-\alpha)\%$ Hedges-type CI as
$[g_{w}-z_{\alpha/2}\widehat{{\rm SE}}_{g_{w}}, g_{w}+z_{\alpha/2}\widehat{{\rm SE}}_{g_{w}}]$,
where 
\begin{equation}\label{se_gw}
	\widehat{{\rm SE}}_{g_{w}}=\widehat{{\rm SE}}_{d_w}\cdot B_{n_1-1,w}B_{n_0-1,1-w}
\end{equation} 
is the estimated standard error of $g_w$
and $\widehat{{\rm SE}}_{d_w}$ is given in (\ref{se_dw}).

%\vspace{0.4cm}
\begin{figure}[H]
	\centering
	\includegraphics[width=1\textwidth]{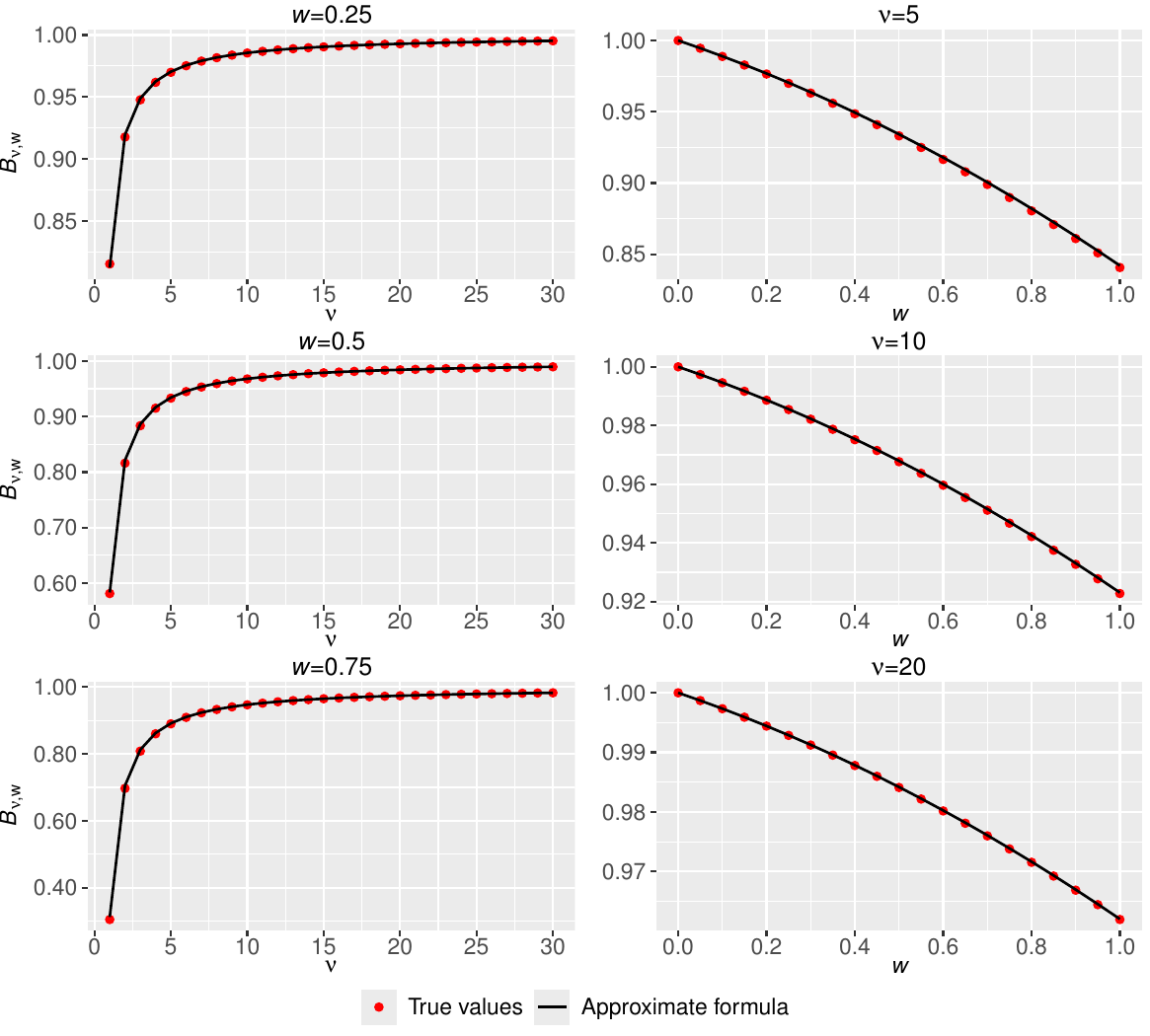}
	\setlength{\abovecaptionskip}{0pt}%
	\setlength{\belowcaptionskip}{0pt}%
	\caption{The red points represent the true values of $B_{\nu,w}$
		and the black curves represent the approximate formulas in
		$B_{\nu,w}$ in (\ref{b}). The left panels are for $w=0.25$, 0.5 and 0.75 with $\nu$ up to 30, and the right panels are for $\nu=5$, 10 and 20 with $w\in[0,1]$.}
	\label{fig:factor1}
\end{figure}

To promote the practical use of (\ref{s}) and (\ref{se_gw}), we further provide a simple approximation 
for the bias correction factor as 
\begin{equation}\label{b}
	B_{\nu,w}\approx 1-\dfrac{(2+w)w}{4\nu-1}.
\end{equation}
When $w=0$, we have $B_{\nu,0}=1$ so that no bias correction is needed  for the related group. 
When $w=1$, the approximate formula reduces to $B_{\nu,1} = J_{\nu} \approx 1-3/(4\nu-1)$ as in \cite{hedges1981distribution}. 
While for other weights, we plot the true values of $B_{\nu,w}$  as well as its approximate
formula on the left panels of Figure \ref{fig:factor1} for $w=0.25$, 0.5 and 0.75
with $\nu$ up to 30, and on the right panels of Figure \ref{fig:factor1} for $\nu=5$, 10 and 20
with $w\in[0,1]$.
By the numerical results, 
it is evident that our 
formula (\ref{b}) provides a very accurate approximation for the bias correction factor under all the  settings, which was also supported by the error analysis as in \cite{hedges1981distribution}. 
Taking $w=0.5$ as an example, 
our approximation has the maximum error within 0.002 as $\nu=1$, 
within 0.0003 as $\nu\geq 10$, and within $3.3\times 10^{-5}$ as $\nu\geq 30$.

To conclude, the Hedges-type estimator further improves the  Cohen-type estimator 
by completely eliminating the bias term. 
Additionally,  by  Gautschi's inequality (\citealp{Gautschi1959}), 
we can derive that $B_{\nu,w}<1$ for all $\nu$ and $w\in(0,1]$. 
Thus by (\ref{se_gw}), it can be seen that the Hedges-type CI is also narrower than the Cohen-type CI.

\section{Simulation studies}\label{simulations}
\noindent
This section carries out simulation studies to
evaluate the numerical performance of our new estimators for the geometric SMD. 
Meanwhile, we also compare them with the classic estimators for the pooled SMD and the arithmetic SMD, respectively. 

\subsection{Comparing with the pooled SMD}\label{comparison_dg}
\noindent
To compare the geometric SMD with the pooled SMD,
our first simulation generates balanced data from $N(2,\sigma_1^2)$ and $N(0,\sigma_0^2)$  for the case and control groups, respectively. 
Without loss of generality, we set $\sigma_0^2=1$ and then  vary $\sigma_1^2$ from $1/16$ to 16 to represent  the degrees of imbalance in the variances.
Or equivalently, we let 
$u=\log_2(\sigma_1^2/\sigma_0^2)$ range from -4 to 4.
For the sample sizes, we consider $n_1=n_0=10$ and $n_1=n_0=50$.
To also explore the weight effect on the geometric SMD, 
we take three different weights as $w=0.25$, 0.5 and 0.75.
Lastly, to compare our new estimators with Cohen's $d$ and Hedges' $g$,  
we numerically calculate the bias, mean squared error (MSE) and coverage probability of the 95\% CI for each method.

%\vspace{0.4cm}
\begin{figure}[H]
	\centering
	\includegraphics[width=1\textwidth]{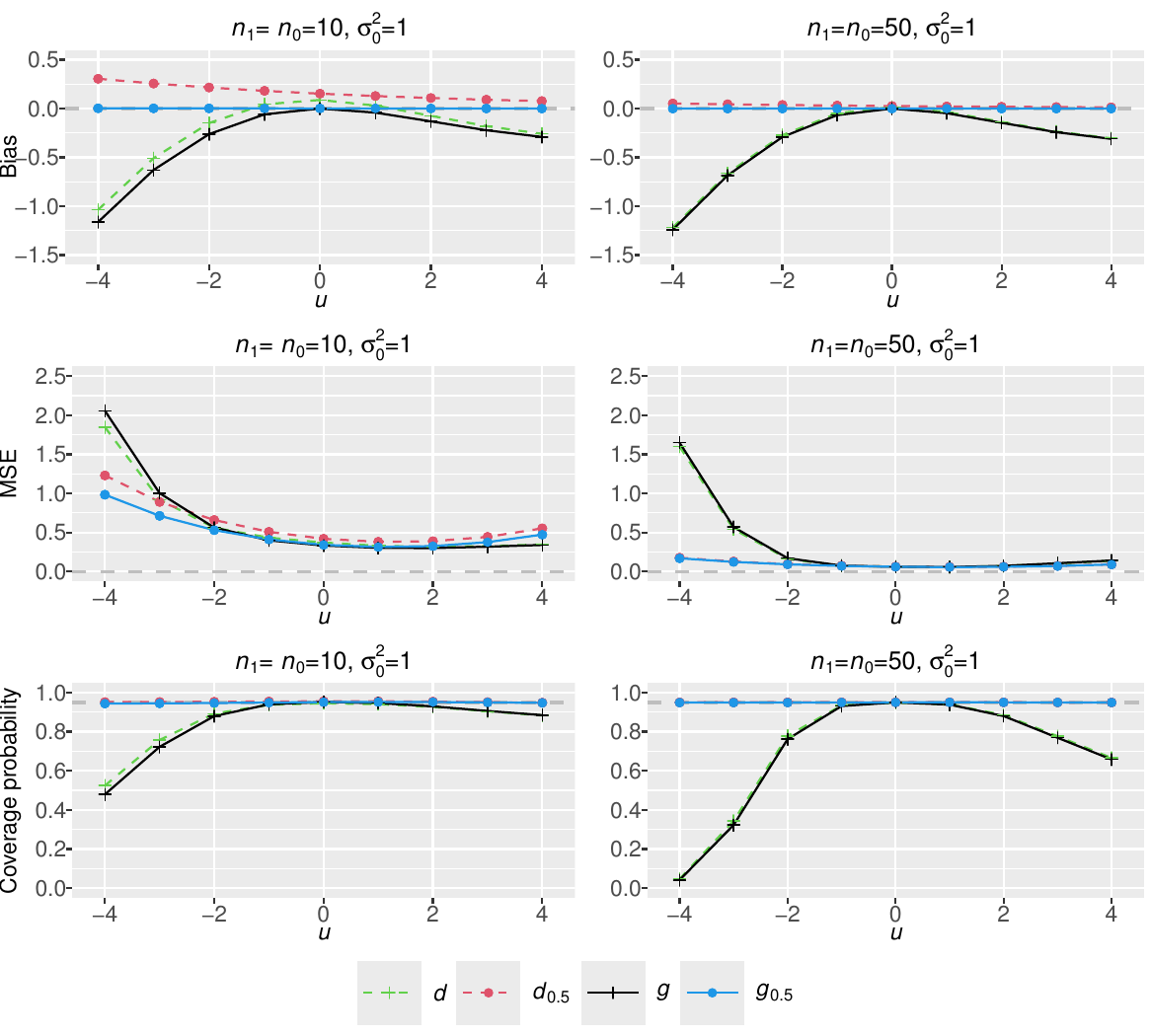}
	\setlength{\abovecaptionskip}{0pt}%
	\setlength{\belowcaptionskip}{0pt}%
	\caption{The bias, MSE and coverage probability for Cohen's $d$, our Cohen-type $d_w$ with $w=0.5$, Hedges' $g$, and our Hedges-type $g_w$ with $w=0.5$ are reported with $\sigma_0^2=1$ and $u$ ranging from -4 to 4, 
		where the left panels are for $n_1=n_0=10$,
		and the right panels are for $n_1=n_0=50$.}
	\label{fig:g_comparison}
\end{figure}

With 1,000,000 simulations, we report the bias, the MSE and the coverage probability for the four estimators in  Figure \ref{fig:g_comparison} for $w=0.5$, and in Appendix E for $w=0.25$ and $w=0.75$.
From the top panels of Figure \ref{fig:g_comparison},
when $u=0$ so that $\sigma_1^2=\sigma_0^2$, 
our Hedges-type estimator in (\ref{s}) performs equally well as Hedges' $g$,
and both of 
them perform better than Cohen's $d$ and our Cohen-type estimator in (\ref{plug-in}). 
On the other hand, when $u\neq 0$ so 
that $\sigma_1^2 \neq \sigma_0^2$,  Cohen's $d$ and Hedges' $g$ will soon get worse if we apply them to estimate
the geometric SMD, since in such cases the pooled SMD cannot be defined. 
In contrast, our Cohen-type and Hedges-type  estimators perform consistently well regardless of  whether the two variances are unbalanced. 
Moreover, by the middle panels of Figure \ref{fig:g_comparison},
our new estimators yield smaller MSE than Cohen's $d$ and Hedges' $g$ in most settings, 
which is

\begin{figure}[H]
	\centering
	\includegraphics[width=1\textwidth]{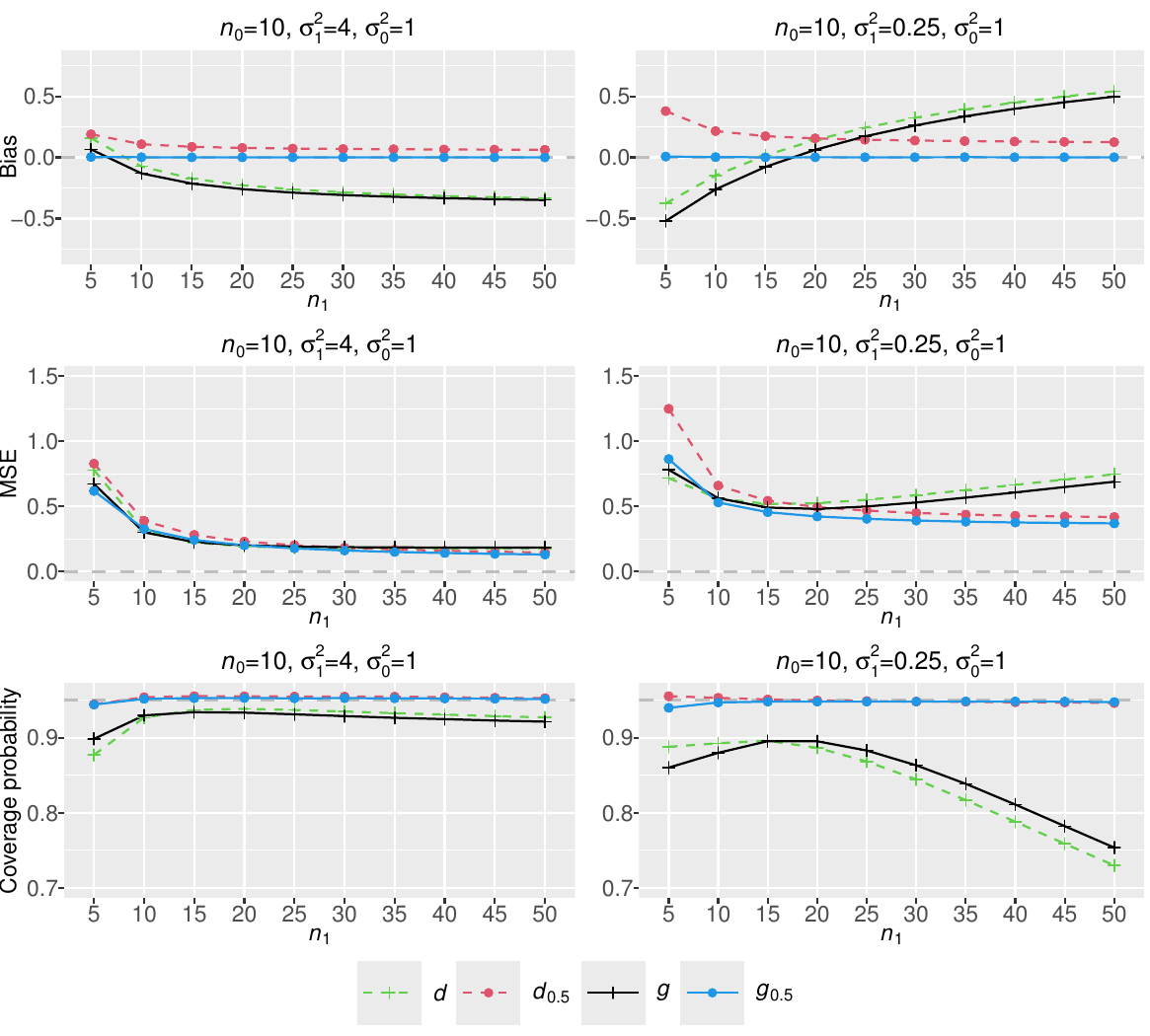}
	\setlength{\abovecaptionskip}{0pt}%
	\setlength{\belowcaptionskip}{0pt}%
	\caption{The bias, MSE and coverage probability for Cohen's $d$, our Cohen-type $d_w$ with $w=0.5$, Hedges' $g$, and our Hedges-type $g_w$ with $w=0.5$ are reported with $n_0=10$ and $n_1$ ranging from 5 to 50, 
		where the left panels are for $\sigma_1^2=4$ and $\sigma_0^2=1$,
		and the right panels are for $\sigma_1^2=0.25$ and $\sigma_0^2=1$.}
	\label{fig:g_unbalanced_comparison}
\end{figure}

\noindent
more evident when  $u$ is small.
In addition, as shown in the bottom panels of Figure \ref{fig:g_comparison},
our Cohen-type and Hedges-type CIs also achieve the nominal level of the coverage probability in most settings, whereas the two classic estimators for the pooled SMD fail to do so when $u$ is away from zero. 
Lastly, it is noteworthy that the above comparison results remain similar also for other weights, as displayed in Figures \ref{fig:geo_pooled_2575} and \ref{fig:geo_pooled_2575_sample} from Appendix E for $w=0.25$ and 0.75, respectively.

To further evaluate their numerical performance, our second simulation generates unbalanced data for the two groups with $n_0=10$ and $n_1$ varying from 5 to 50 to represent the degrees of imbalance in the sample sizes. 
For the variances, we consider two combinations as  $(\sigma_1^2,\sigma_0^2)=(4,1)$
and $(\sigma_1^2,\sigma_0^2)=(0.25,1)$.
All other settings are kept the same as in the first simulation including the number of simulations. 
We then report the bias, the MSE and the coverage probability for $w=0.5$ in Figure \ref{fig:g_unbalanced_comparison}; and to save space, the simulation  results for other weights will be omitted since they are similar to those for $w=0.5$. 
From the top panels of Figure \ref{fig:g_unbalanced_comparison}, 
our Hedges-type estimator is, again, nearly unbiased in all settings.
By the middle panels, it is also seen that our Hedges-type estimator provides the smallest MSE in most settings compared with the other methods. 
Lastly, the bottom panels show that  the coverage probabilities of the 95\% CIs for Cohen's $d$ and Hedges' $g$ are consistently lower than the nominal level.
While for our new CIs, both of them yield a coverage probability very close
to 95\% even if the data are largely unbalanced.

\subsection{Comparing with the arithmetic SMD}\label{comparison_ari}
\noindent
As seen in Section \ref{comparison_dg}, Cohen's $d$ and Hedges' $g$ only perform well when the two variances are equal. 
To further demonstrate the superiority of our new SMD that allows for unequal variances, we now conduct additional simulation studies  to compare it with the arithmetic SMD through their respective estimators.

Specifically,  our third simulation generates two groups of data with sizes $n_1$ and $n_0$ from $N(2,\sigma_1^2)$ and $N(0,\sigma_0^2)$, respectively. 
We also consider  two distinct but complementary settings,
where one is with balanced sample sizes but unequal variances 
($n_1=n_0=10$, $\sigma_1^2=4$, $\sigma_0^2=1$),
and the other is with unbalanced sample sizes but equal variances
($n_1=30$, $n_0=10$, $\sigma_1^2=\sigma_0^2=1$).
Then for each $w\in[0,1]$, we apply the estimators $d_w^*$ in (\ref{dw*}) and $g_w^*$ in (\ref{g_w*}) to estimate the arithmetic SMD, and the estimators in (\ref{plug-in}) and (\ref{s}) to estimate the geometric SMD, followed by their respective 95\% CIs as specified. 
Lastly, for a fair comparison between the geometric and arithmetic SMD estimators,  we numerically compute
the relative bias of each method as
\begin{equation*}
	\dfrac{1}{T}\cdot \sum_{i=1}^T \dfrac{{\rm the\ estimated\ SMD}_i-{\rm the\ true\ SMD}}{{\rm the\ true\ SMD}},
\end{equation*}
together with the relative MSE as
\begin{equation*}
	\dfrac{1}{T}\cdot \sum_{i=1}^T \left(\dfrac{{\rm the\ estimated\ SMD}_i-{\rm the\ true\ SMD}}{{\rm the\ true\ SMD}}\right)^2,
\end{equation*}
where $T=1,000,000$ is the number of simulations for each setting.

\begin{figure}[H]
	\centering
	\includegraphics[width=1\textwidth]{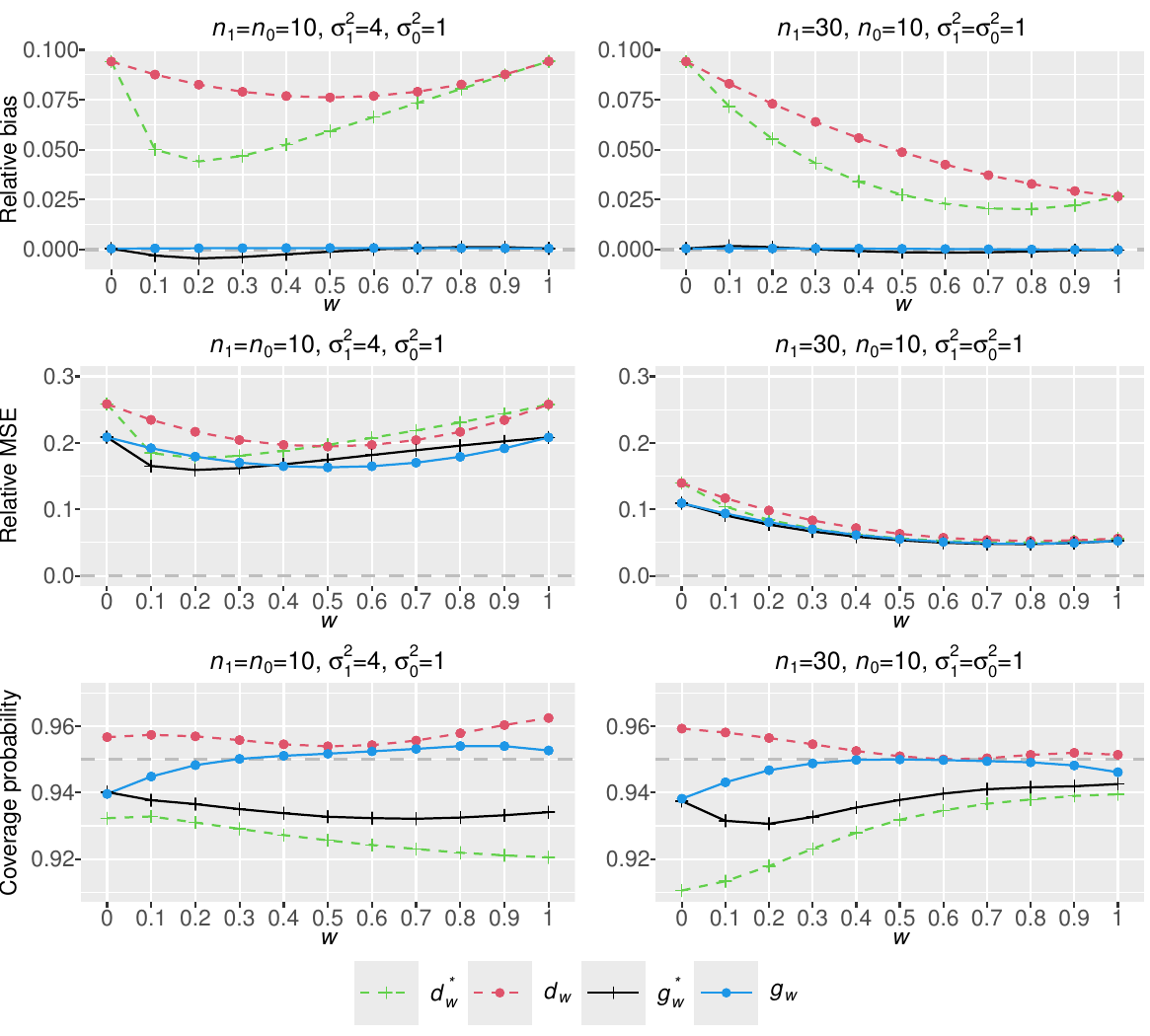}
	\setlength{\abovecaptionskip}{0pt}%
	\setlength{\belowcaptionskip}{0pt}%
	\caption{The relative bias, relative MSE, coverage probability for arithmetic Cohen-type $d_{w}^*$, our Cohen-type $d_{w}$, arithmetic Hedges-type $g_{w}^*$ and our Hedges-type $g_w$
		with $w\in[0,1]$ are reported, where the left panels are for balanced sample sizes but unequal variances, and the right panels are for unbalanced sample sizes but equal variances.}
	\label{fig:d2_comparison}
\end{figure}

We report the relative bias, the relative MSE, and the coverage probability for each of the four estimators in  Figure \ref{fig:d2_comparison}. 
The top panels show that the two Cohen-type estimators both result in  a significant bias.
Additionally, we note that the arithmetic Hedges-type estimator also yields a non-negligible bias under both scenarios,
indicating that the bias correction in $g_w^*$ is incomplete.
%In addition, the bias of $g_w^*$ is also related to the weight $w$ as well as the settings of variances and sample sizes.
%Specifically from the top left panel of Figure \ref{fig:d2_comparison}, $g_{0.5}^*$ yields the obviously larger bias compared to $g_w^*$ with other weights.
This, in fact, coincides with our derivation in (\ref{huynh}) that  
$d_{w}^*$ does not exactly follow the noncentral $t$-distribution with the degrees of freedom as $\nu_{w}=[(w\sigma_1^2+(1-w)\sigma_0^2)^2]/[w^2\sigma_1^4/(n_1-1)+(1-w)^2\sigma_0^4/(n_0-1)]$,
and subsequently the bias correction term $J_{\nu_w}$ cannot fully eliminate the bias.
Moreover, with $\nu_w$ unknown in practice, one needs to further apply the plug-in estimate
$\hat{\nu}_{w}=[(wS_1^2+(1-w)S_0^2)^2]/[w^2S_1^4/(n_1-1)+(1-w)^2S_0^4/(n_0-1)]$, which also introduces new errors.
On the other hand, with the geometric weighting approach, our new estimator $g_w$
is able to completely eliminate the bias associated with the Cohen-type estimator under any combination of the weight, the variances, and the sample sizes.
Note that the superiority of our geometric Hedges-type estimator is also evidenced by the middle panels of Figure \ref{fig:d2_comparison} from the perspective of the relative MSE.
From the bottom panels of Figure \ref{fig:d2_comparison}, it is clear that the two CIs for the arithmetic SMD are not able to give 95\% 

\vspace{0.4cm}

\noindent
coverage probability. 
In contrast, both of our Cohen-type and Hedges-type CIs can  provide a coverage probability very close to  the nominal level, especially when the weight is in  the middle of $[0,1]$. 
Lastly,  following the discussion at the end of Section \ref{hedges_type_w}, our Hedges-type CI is narrower than our Cohen-type CI, which explains why its coverage probability is also slightly lower than that of its counterpart.

To sum up, when the assumption of equal variance is violated,
the pooled SMD in (\ref{smd_equal_variance}) becomes invalid  so that the well-known Cohen's $d$ and Hedges' $g$ can also be misleading.
Furthermore, the arithmetic SMD has its limitations in the sense that an unbiased estimate is hard to obtain, in addition to its complex form which may also hinder its real application. 
As an alternative, our newly introduced geometric SMD, along with  the Cohen-type and Hedges-type estimators, bring new insights to the existing literature and can thus be  highly recommended for practical applications.

\section{Application to meta-analysis}\label{real_data}
\noindent
To illustrate the usefulness of the new SMD and its estimation,
we present a real data example in meta-analysis.
\cite{GS2023} carried out a systematic review and several meta-analyses to evaluate the effect of  the cognitive stimulation (CS) on psychosocial results in older adults.
They proposed to demonstrate that the quality of life (QoL), depression, anxiety and activities of daily living can be improved independent of the pharmacological treatment such as acetylcholinesterase inhibitors by CS.

\begin{figure}[H]
	\centering
	\includegraphics[width=1.3\textwidth,angle=90]{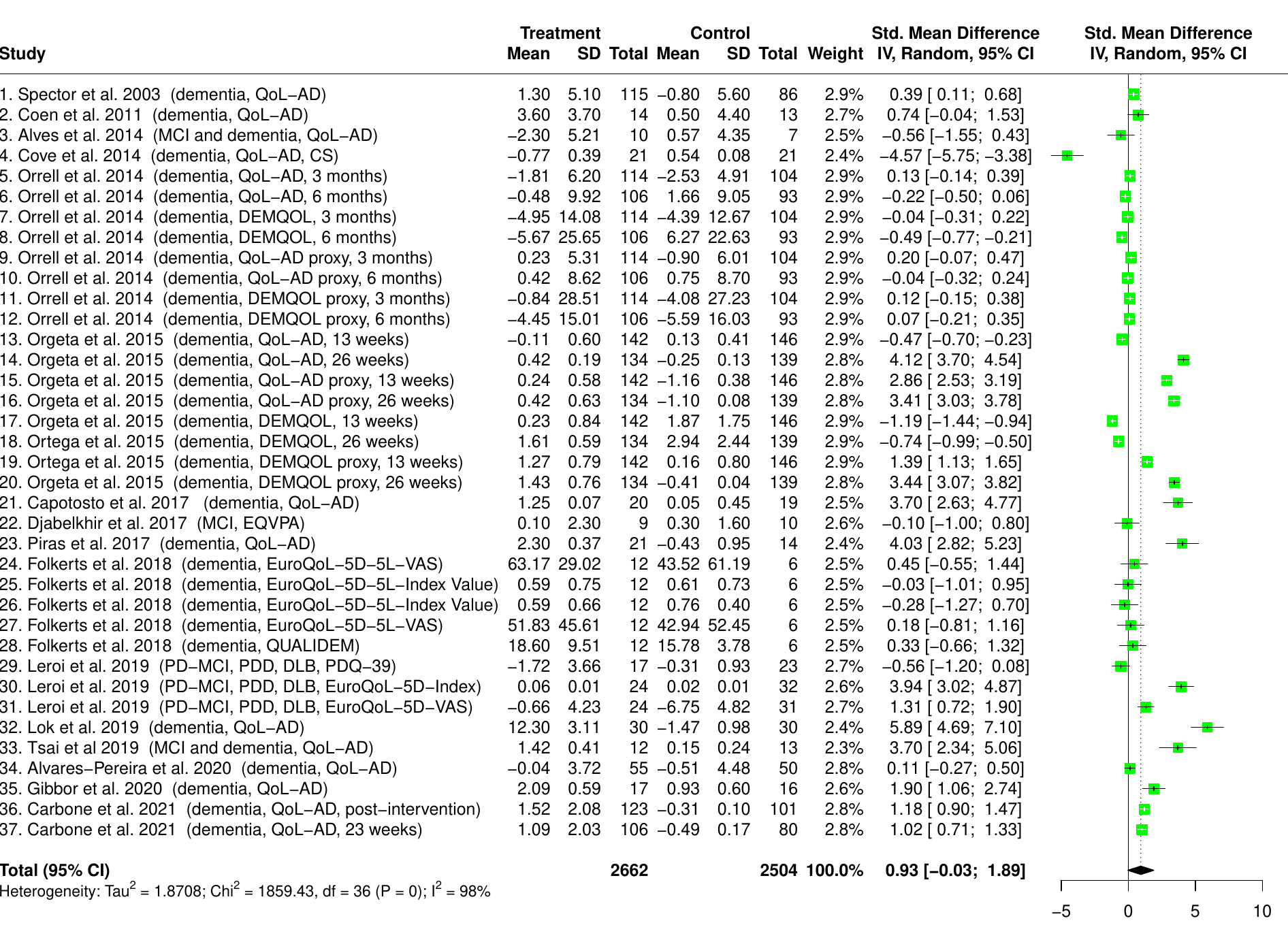}
	\setlength{\abovecaptionskip}{0pt}%
	\setlength{\belowcaptionskip}{0pt}%
	\caption{The original meta-analysis results in \cite{GS2023}.}
	\label{fig:original_results}
\end{figure}

In this section, we focus on the meta-analysis for QoL
and report its original results in Figure \ref{fig:original_results}.
\cite{GS2023} used the SMD to measure the effect in each study 
and applied Hedges' $g$ to estimate the SMD.
To accommodate studies with data collected at different follow-up times, 
the meta-analysis employed the robust variance estimator by \cite{Tipton2015}  and applied the Satterthwaite adjustment to the degrees of freedom in the random-effects model (\citealp{Borenstein2009}).
It is noted from the forest plot in Figure \ref{fig:original_results}
that the overall SMD estimate is 0.93 with the 95\% CI
being $[-0.03,1.89]$, which shows an insignificant result in favor of the case group.
The lower bound of the 95\% CI is very close to 0 and the $p$-value is 0.056, 
which causes difficulty in making a convincing conclusion.
We note that the unequal variances appear in many studies included
in the meta-analysis.
Taking the 20th study as an example, the sample standard deviation from the case group is
0.76, which is 19 times of the sample standard deviation 0.04 from the control group.
Moreover, there are 13 (out of a total of 37) studies with the ratios of sample standard deviations between the case 
and control groups outside

\begin{figure}[H]
	\centering
	\includegraphics[width=1.3\textwidth,angle=90]{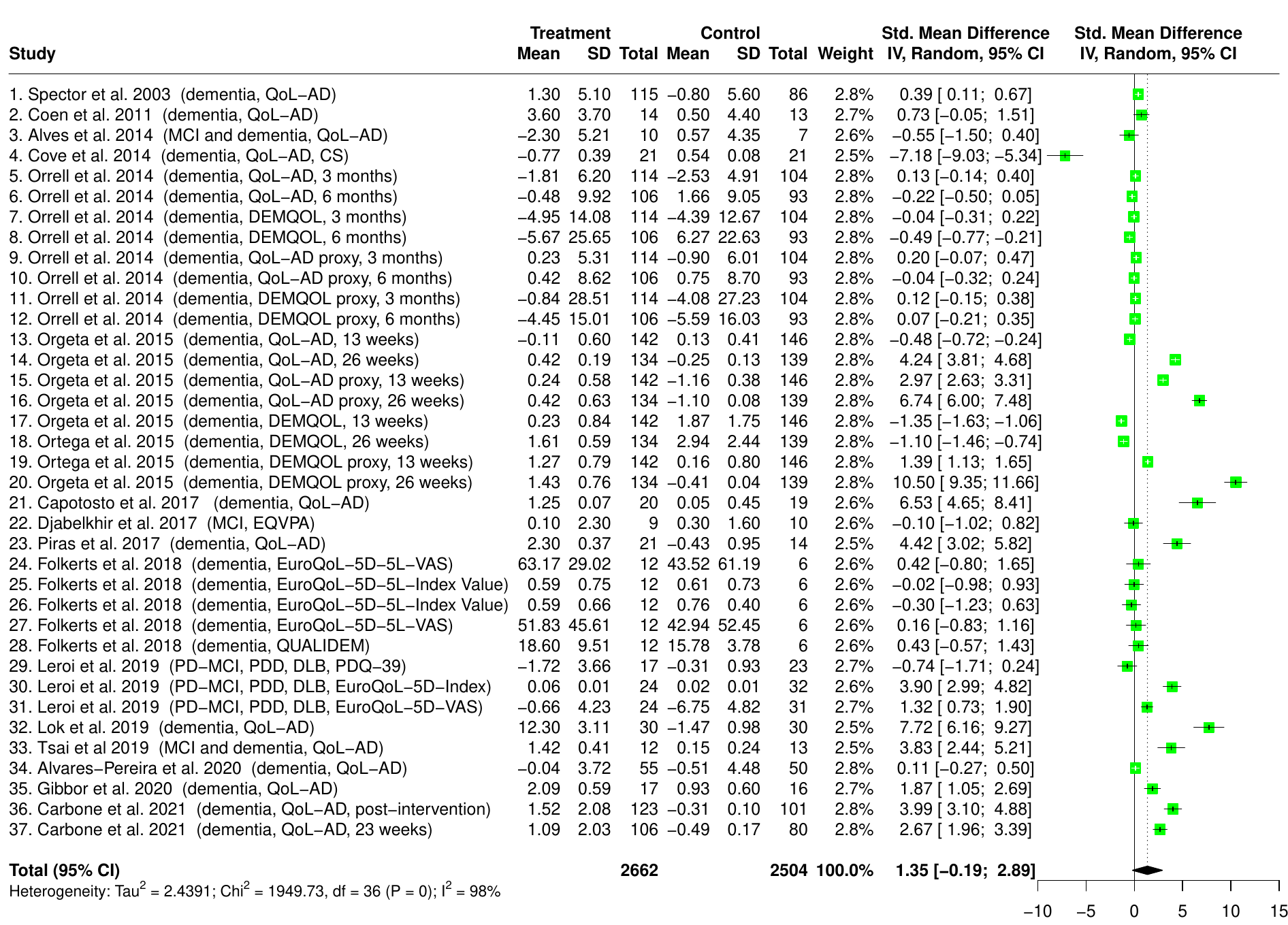}
	\setlength{\abovecaptionskip}{0pt}%
	\setlength{\belowcaptionskip}{0pt}%
	\caption{The new results by our geometric SMD and its Hedges-type estimation.}
	\label{fig:our_results}
\end{figure}

\noindent
the interval  $[0.5, 2]$. 
This indicates that  the pooled SMD in (\ref{smd_equal_variance}) may no longer be appropriate  as the  true measure, and so is  Hedges' $g$ 
to estimate the SMD.

To handle the unequal variances, we apply both the arithmetic and geometric SMDs with $w=0.5$, as well as  their Hedges-type estimates, for all 37 studies included in the meta-analysis. 
For the arithmetic SMD, we note that the SMD estimates for the individual studies
do not differ much from those from Hedges' $g$.
This is mainly because of the approximate equivalence between $g$ and $g_{0.5}^*$
for the studies with nearly balanced sample sizes,
regardless of whether or not  the sample standard deviations largely differ. 
As a consequence, the meta-analytical results remain similar
with the overall SMD estimate being 0.9 and the 95\% CI being $[-0.04,1.83]$.
Moreover, the $p$-value for the overall SMD estimate is 0.057, 
which also suffers from the difficulty in making a convincing conclusion.
For more details, please refer to the forest plot in Figure \ref{fig:arithmetic_results} of Appendix F.

In contrast, our geometric SMD significantly alters the individual SMD estimates, which, in turn, has a notable impact on the overall  meta-analytical results.
From Figure \ref{fig:our_results} for our new results, we observe that the SMD estimate in the 20th study
is 10.5 with the 95\% CI being $[9.35,11.66]$
while the original results in Figure \ref{fig:original_results}
show the SMD estimate as 3.44 with 95\% CI being $[3.07,3.82]$.
This also occurs similarly  for other studies with unequal sample standard deviations,
e.g., studies 4, 16, 21 and so on.
Consequently, the meta-analysis leads to the overall SMD estimate as 1.35 with the 95\% CI being $[-0.19,2.89]$,
and the $p$-value is 0.076.
Based on the new results, it is more convincing to draw the conclusion of an insignificant result in favor of the case group.
Taken together, it thus offers fresh perspectives for our new method to serve as a valuable  alternative to
Hedges' $g$ for practical use.

\section{Conclusions}\label{discussions}
\noindent
For clinical studies with continuous outcomes, the standardized mean difference
(SMD) is the most widely used effect size
that can measure the raw data of different studies on to the same scale.
Under the equal variance assumption, the pooled SMD is clearly defined as
the mean difference divided by the common standard deviation,
together with Cohen's $d$ and Hedges' $g$ as the popular estimates. 
In many clinical studies, however, the equal variance assumption may not be realistic,
presenting a significant limitation for the practical use of Cohen's $d$ and Hedges' $g$ (\citealp{hedges2025}).
To deal with the unequal variances, 
several definitions for standardizing the mean difference had been introduced in the previous literature, 
with the essential idea being to take the arithmetic mean of the two variances for standardization, referred to as the arithmetic SMD. 
Nevertheless, as has been extensively discussed, this measure has serious limitations including, but not limited to, the complex estimation form, unavoidable bias, and low coverage probability. 
These limitations restrict its practical utility, 
leaving Cohen's $d$ and Hedges' $g$ as the default choices 
even with largely unbalanced variances.

One main contribution of this paper is to introduce a new definition of SMD as $\delta_w=(\mu_1-\mu_0)/(\sigma_1^w\sigma_0^{1-w})$
by taking the geometric mean of the two variances for standardization. 
To the best of our limited knowledge, this is the first work to consider the  geometric weighting for the standardizer in (\ref{smd_general}), which makes our new SMD very unique compared to  the conventional arithmetic SMD.
We further develop the easy-to-implement estimators of the geometric SMD using the Cohen-type and Hedges-type methods. 
Notably, the geometric weighting approach enables an unbiased Hedges-type estimator, which is a critical advantage unattainable with the arithmetic SMD.
Simulation results show that our new estimators are  superior to the existing ones under a wide range of settings in terms of the relative bias, relative MSE, and coverage probability. 
We also apply our new SMD to a real-world meta-analysis
involving many studies with unequal variances,
which results in more convincing results compared to the original results with Hedges' $g$ as the effect size. 
To conclude,  our geometric SMD does not impose  the equal variance assumption
as in the pooled SMD, and moreover it has some nice properties and advantages that the arithmetic SMD does not have.
And accordingly, as the effect size for meta-analysis,
our Cohen-type and Hedges-type estimators for the new SMD can also be highly recommended to supersede the original Cohen's $d$ and Hedges' $g$.

\newpage

\setcounter{equation}{0}
\renewcommand{\theequation}{S\arabic{equation}}

\begin{center}
	 \textbf{\large Appendix A: Derivation of formula (\ref{huynh})}
\end{center}
 
\noindent
%\noindent{\bf Derivations of formula (\ref{huynh}).}
The Cohen-type estimator of $\delta_w^*$ is given as $d_w^*=(\bar{X}_1-\bar{X}_0)/\sqrt{wS_1^2+(1-w)S_0^2}$.
It is evident that $\bar{X}_1-\bar{X}_0\sim N(\mu_1-\mu_0,\sigma_1^2/n_1+\sigma_0^2/n_0)$,
and that the numerator and denominator are mutually independent.
Furthermore, according to the Welch-Satterthwaite equation, it approximately holds for any  $w\in(0,1)$ that (\citealp{welch1938significance,Satterthwaite1941})
\begin{equation*}
	\dfrac{\nu(wS_1^2+(1-w)S_0^2)}{w\sigma_1^2+(1-w)\sigma_0^2}\sim \chi^2_\nu,
\end{equation*}
where 
$$\nu=\dfrac{(w\sigma_1^2+(1-w)\sigma_0^2)^2}{w^2\sigma_1^4/(n_1-1)+(1-w)^2\sigma_0^4/(n_0-1)}.$$
Taken together, it approximately holds that
\begin{equation*}
	\left(\dfrac{\sigma_1^2/n_1+\sigma_0^2/n_0}{w\sigma_1^2+(1-w)\sigma_0^2}\right)^{-1/2}\cdot \dfrac{\bar{X}_1-\bar{X}_0}{\sqrt{wS_1^2+(1-w)S_0^2}}\sim t(\nu,\lambda),
\end{equation*}
where $\lambda=(\mu_1-\mu_0)/\sqrt{\sigma_0^2/n_0+\sigma_1^2/n_1}$.
\\

\begin{center}
	\textbf{\large Appendix B: Theorem \ref{theorem1} and its proof} 
\end{center}
\begin{theorem}\label{theorem1}
	Let $c=\lim\limits_{n_1,n_0\rightarrow\infty}n_1/n_0$. Then for the Cohen-type estimator $d_w$ in (13), we have 
	\begin{equation*}
		\left(\dfrac{\sigma_1^2}{n_1}+\dfrac{\sigma_0^2}{n_0}\right)^{-1/2} (d_{w}-\delta_{w})  \overset{D}\rightarrow N\left(0,\dfrac{\delta_{w}^2}{2}\cdot \dfrac{w^2+c(1-w)^2}{\sigma_1^2+c\sigma_0^2} +\dfrac{1}{\sigma_1^{2w}\sigma_0^{2-2w}}\right)
	\end{equation*}
	as $n_1, n_0 \to \infty$, where $\overset{D}\rightarrow$ denotes the convergence in distribution.
\end{theorem}	

{\bf Proof.}
First, as $n_i$ tends to infinity for $i=0$ and 1, by the asymptotic normality of the sample variance, we have
\begin{equation*}\label{S_asymptotic}
	\sqrt{n_i}(S_i^2-\sigma_i^2)\overset{D}\rightarrow N(0,2\sigma_i^4).
\end{equation*}
By the delta method, we have
\begin{equation}\label{1S_asymptotic}
	\sqrt{n_i}(S_i^{-w}-\sigma_i^{-w})\overset{D}\rightarrow N(0,w^2/(2\sigma_i^{2w})).
\end{equation}
Then by (\ref{1S_asymptotic}), as $n_1$ and $n_0$ tend to infinity, 
we have $S_1^{-w}S_0^{w-1}\overset{P}\rightarrow \sigma_1^{-w}\sigma_0^{w-1}$,
where $\overset{P}\rightarrow$ represents the convergence in probability.
In addition, as $n_1$ and $n_0$ tend to infinity, we have 
\begin{equation*}
	\sqrt{\left(\dfrac{\sigma_1^2}{n_1}+\dfrac{\sigma_0^2}{n_0}\right)^{-1}}(\bar{X}_1-\bar{X}_0-(\mu_1-\mu_0))\overset{D}\rightarrow N(0,1).
\end{equation*}
Furthermore, we have
\begin{equation*}
	\sqrt{\left(\dfrac{\sigma_1^2}{n_1}+\dfrac{\sigma_0^2}{n_0}\right)^{-1}} (d_{w}-\delta_{w})  = {\rm I}+{\rm II}+{\rm III}+{\rm IV}+{\rm V},\\
\end{equation*}
where
\begin{align*}
	{\rm I}
	& = \sqrt{\left(\dfrac{\sigma_1^2}{n_1}+\dfrac{\sigma_0^2}{n_0}\right)^{-1}}(\bar{X}_1-\bar{X}_0-(\mu_1-\mu_0))\cdot (S_1^{-w}S_0^{w-1}- \sigma_1^{-w}\sigma_0^{w-1})\\
	{\rm II}& = \sqrt{\left(\dfrac{\sigma_1^2}{n_1}+\dfrac{\sigma_0^2}{n_0}\right)^{-1}}(\bar{X}_1-\bar{X}_0-(\mu_1-\mu_0))\cdot \sigma_1^{-w}\sigma_0^{w-1}\\
	{\rm III}& = \sqrt{\left(\dfrac{\sigma_1^2}{n_1}+\dfrac{\sigma_0^2}{n_0}\right)^{-1}}(\mu_1-\mu_0)(S_1^{-w}-\sigma_1^{-w})(S_0^{w-1}-\sigma_0^{w-1})\\
	{\rm IV}& = \sqrt{\left(\dfrac{\sigma_1^2}{n_1}+\dfrac{\sigma_0^2}{n_0}\right)^{-1}}\dfrac{\mu_1-\mu_0}{\sigma_1^w}\cdot	(S_0^{w-1}-\sigma_0^{w-1})\\
	{\rm V}& = \sqrt{\left(\dfrac{\sigma_1^2}{n_1}+\dfrac{\sigma_0^2}{n_0}\right)^{-1}}\dfrac{\mu_1-\mu_0}{\sigma_0^{1-w}}\cdot 	(S_1^{-w}-\sigma_1^{-w}).
\end{align*}
As $n_1$ and $n_0$ tend to infinity, we have
${\rm I}\overset{P}\rightarrow 0, {\rm II}\overset{D}\rightarrow N(0,\sigma_1^{-2w}\sigma_0^{2w-2}),$ and ${\rm III}\overset{P}\rightarrow 0.$
For IV and V, we need to further clarify the order of $n_1/n_0$, and
specifically, let $c=\lim\limits_{n_1,n_0\rightarrow\infty}n_1/n_0$.

\begin{itemize}
	\item As $c=0$, ${\rm IV}\overset{P}\rightarrow 0$ and ${\rm V}\overset{D}\rightarrow N(0,w^2\delta_{w}^2/(2\sigma_1^2))$.
	\item As $c$ is fixed and positive, 
	${\rm IV}\overset{D}\rightarrow N(0,c(1-w)^2\delta_{w}^2/[2(\sigma_1^2+c\sigma_0^2)])$ and ${\rm V}\overset{D}\rightarrow N(0,w^2\delta_{w}^2/[2(\sigma_1^2+c\sigma_0^2)])$.
	\item As $c\rightarrow \infty$, ${\rm IV}\overset{D}\rightarrow N(0,(1-w)^2\delta_{w}^2/(2\sigma_0^2))$ and ${\rm V}\overset{P}\rightarrow 0$.
\end{itemize}
Finally, by the mutual independence among $\bar{X}_1$, $\bar{X}_0$, $S_1$ and $S_0$, and Slutsky's Theorem,
we conclude that
\begin{equation*}
	\left(\dfrac{\sigma_1^2}{n_1}+\dfrac{\sigma_0^2}{n_0}\right)^{-1/2} (d_{w}-\delta_{w})  \overset{D}\rightarrow N\left(0,\dfrac{\delta_w^2}{2}\cdot \dfrac{w^2+c(1-w)^2}{\sigma_1^2+c\sigma_0^2}+\dfrac{1}{\sigma_1^{2w}\sigma_0^{2-2w}}\right).
\end{equation*}
Thus as $n_1$ and $n_0$ tend to infinity,
$d_w$ follows a normal distribution with mean $\delta_w$
and variance as
\begin{equation*}\label{var_dw}
	{\rm Var}(d_w)= \dfrac{\delta_{w}^2}{2} \left(\dfrac{w^2}{n_1}+\dfrac{(1-w)^2}{n_0}\right) +\dfrac{n_1(\sigma_0^2/\sigma_1^2)^w+n_0(\sigma_1^{2}/\sigma_0^2)^{1-w}}{n_1n_0}.
\end{equation*}

\

\begin{center}
	\textbf{\large Appendix C: Theorem \ref{theorem2} and its proof} 
\end{center}
\begin{theorem}\label{theorem2}
	Let $B_{\nu,w}=(2/\nu)^{w/2} \Gamma(\nu/2)/\Gamma((\nu-w)/2)$.
	Then for the Cohen-type estimator $d_{w}$ in (13), we have
	$$ E(d_{w})=
	\dfrac{\delta_{w}}{B_{n_1-1,w}B_{n_0-1,1-w}}$$ 
	and 
	\begin{equation}\label{Vard}
		{\rm Var}(d_{w}) 
		=\dfrac{\delta_{w}^2}{B_{n_1-1,2w}B_{n_0-1,2-2w}}-\dfrac{\delta_{w}^2}{(B_{n_1-1,w}B_{n_0-1,1-w})^2}
		+\dfrac{(\sigma_1/\sigma_0)^{2-2w}/n_1+(\sigma_0/\sigma_1)^{2w}/n_0}{B_{n_1-1,2w}B_{n_0-1,2-2w}}.
	\end{equation}
\end{theorem}

{\bf Proof.}
Note that $\bar{X}_1$, $\bar{X}_0$, $S_1^2$ and $S_0^2$ are independent, thus 
\begin{equation*}
	E(d_{w})=
	(\mu_1-\mu_0)E\left(\frac{1}{S_1^w}\right)  E\left(\frac{1}{S_0^{1-w}}\right). 
	\label{Ed}
\end{equation*} 
Denote $Y_1=\sigma_1^2/[(n_1-1)S_1^2] \sim   IG((n_1-1)/2,1/2)$,
where $IG(\alpha,\beta)$ indicates the inverse-gamma distribution. 
Then we have $1/S_1^w=[(n_1-1)Y_1/\sigma_1^2]^{w/2}$, and for $n_1>1$,
\begin{equation*}
	E\left(\dfrac{1}{S_1^w}\right)=\frac{1}{\sigma_1^w}\left(\frac{n_1-1}{    2  }\right)^{ w/2} \frac{\Gamma((n_1-1-w)/2) }{\Gamma({(n_1-1)/2})}.
\end{equation*}
By denoting $B_{\nu,w}=(2/\nu)^{w/2} \Gamma(\nu/2)/\Gamma((\nu-w)/2)$, 
we have $E(1/S_1^w)=1/(B_{\nu_1,w}\sigma_1^w)$, 
and similarly, $E(1/S_0^{1-w})=1/(B_{\nu_0,1-w}\sigma_0^{1-w})$,
where $\nu_1=n_1-1$ and $\nu_0=n_0-1$.
Thus we have
\begin{equation*}
	E(d_{w}) = \dfrac{\delta_{w}}{B_{\nu_1,w}B_{\nu_0,1-w}}.
	\label{Ed2}
\end{equation*} 

Next, we calculate the variance of $d_{w}$ as
	\begin{align*}	
		{\rm Var}(d_{w})&=E\left[\frac{(\bar{X}_1-\bar{X}_0)^2}{   {S}_1^{2w}  {S}_0^{2-2w} }\right]-\left[E\left(\frac{\bar{X}_1-\bar{X}_0}{   {S}_1^w  {S}_0^{1-w} }\right)\right]^2\nonumber \\
		&=E[(\bar{X}_1-\bar{X}_0)^2]E\left(\frac{1}{  {S}_1^{2w} }\right)  E\left(\frac{1}{  {S}_0^{2-2w} }\right) -\left[E\left(\frac{\bar{X}_1-\bar{X}_0}{   {S}_1^w  {S}_0^{1-w} }\right)\right]^2\nonumber\\
		&=\dfrac{(\mu_1-\mu_0)^2+\sigma_1^2/n_1 +\sigma_0^2/n_0}  
		{\sigma_1^{2w} \sigma_0^{2-2w}B_{\nu_1,2w}B_{\nu_0,2-2w}}-\left(\frac{ \mu_1-\mu_0 }{\sigma_1^w \sigma_0^{1-w}}\right)^2 \left(\dfrac{1}{B_{\nu_1,w}B_{\nu_0,1-w}}\right)^2 \nonumber  \\
		&= \left(\dfrac{1}{B_{\nu_1,2w}B_{\nu_0,2-2w}}-\dfrac{1}{(B_{\nu_1,w}B_{\nu_0,1-w})^2}\right) \delta_{w}^2
		+\dfrac{1}{B_{\nu_1,2w}B_{\nu_0,2-2w}}\left(\dfrac{\sigma_1^{2-2w}}{n_1\sigma_0^{2-2w}}+\dfrac{\sigma_0^{2w}}{n_0\sigma_1^{2w}}\right).
	\end{align*}

\

\begin{center}
	\textbf{\large Appendix D: Corollary \ref{corollary} and its proof} 
\end{center}
\begin{corollary}\label{corollary}
	The exact variance of $d_w$ in (\ref{Vard}) can be approximated as 
	\begin{equation*}
		{\rm Var}(d_w)\approx \dfrac{\delta_{w}^2}{2}\left(\dfrac{w^2}{n_1-1}+\dfrac{(1-w)^2}{n_0-1}\right) +\dfrac{(n_1-1)(\sigma_0^2/\sigma_1^2)^w+(n_0-1)(\sigma_1^{2}/\sigma_0^2)^{1-w}}{(n_1-1)(n_0-1)}.
	\end{equation*}
\end{corollary}

{\bf Proof.}
Note that the exact variance of $d_w$ in (\ref{Vard}) is in quite a complicated format 
and not ready to be used in practice.
We thus carry out some approximations as follows.
By the approximation $B_{\nu,w}\approx 1-(2+w)w/(4\nu-1)$, we have
\begin{align*}\label{approximated_variance}
	\dfrac{1}{B_{\nu_1,2w}B_{\nu_0,2-2w}}-\dfrac{1}{(B_{\nu_1,w}B_{\nu_0,1-w})^2}
	=&\
	\dfrac{(B_{\nu_1,w}B_{\nu_0,1-w})^2-B_{\nu_1,2w}B_{\nu_0,2-2w}}{B_{\nu_1,2w}B_{\nu_0,2-2w}(B_{\nu_1,w}B_{\nu_0,1-w})^2} \nonumber\\
	=&\ \dfrac{128w^2\nu_1\nu_0^2+128(1-w)^2\nu_0\nu_1^2+R_1}{256\nu_1^2\nu_0^2+R_2},
\end{align*}
where $R_1$ and $R_2$ indicate the remaining terms given as
\begin{align*}
	R_1 &= 
	8(2x^2\nu_0^2+2y^2\nu_1^2+8xy\nu_1\nu_0+x^2y\nu_0+xy^2\nu_1)+x^2y^2
	-4(\nu_1+\nu_0)(4a\nu_0+4b\nu_1+ab),	\\
	R_2 &= [(8a\nu_1+a^2)(4\nu_0+b)^2+(8b\nu_0+b^2)(4\nu_1+a)^2]B_{\nu_1,2w}B_{\nu_0,2-2w},
\end{align*}
where $x=-1-(2+w)w$, $y=-1-(3-w)(1-w)$, $a=-1-2w(2+2w)$ and $b=-1-(4-2w)(2-2w)$.
When the sample sizes are not very small, noting that the remaining terms
$R_1$ and $R_2$ are much smaller than the respective dominating terms,
it approximately holds that 
\begin{equation*}\label{approximated_variance}
	\dfrac{1}{B_{\nu_1,2w}B_{\nu_0,2-2w}}-\dfrac{1}{(B_{\nu_1,w}B_{\nu_0,1-w})^2}
	\approx \dfrac{w^2}{2(n_1-1)}+\dfrac{(1-w)^2}{2(n_0-1)}.
\end{equation*}

For $(B_{\nu_1,2w}B_{\nu_0,2-2w})^{-1}$ in the second term of (\ref{Vard}),
we note that it is slightly larger than 1 for not very small sample sizes and approaches to 1 as $\nu_1$ and $\nu_0$ increase, 
which can also be demonstrated by Figure \ref{fig:factor1}.
By balancing the accuracy and simplicity, the second term in (\ref{Vard}) can be approximated as
\begin{equation*}
	\dfrac{1}{B_{\nu_1,2w}B_{\nu_0,2-2w}}\left(\dfrac{\sigma_1^{2-2w}}{n_1\sigma_0^{2-2w}}+\dfrac{\sigma_0^{2w}}{n_0\sigma_1^{2w}}\right)\approx \dfrac{\sigma_1^{2-2w}}{(n_1-1)\sigma_0^{2-2w}}+\dfrac{\sigma_0^{2w}}{(n_0-1)\sigma_1^{2w}}.
\end{equation*}
Therefore, the large sample variance in (\ref{var_dw1}) can be, in fact, adjusted to approximate the exact variance 
if we replace $n_1$ and $n_0$ by $n_1-1$ and $n_0-1$, respectively.

\

\setcounter{figure}{0}
\renewcommand{\thefigure}{S\arabic{figure}}

\begin{center}
	\textbf{\large Appendix E: Supplementary figures for Section \ref{comparison_dg}} 
\end{center}

\vspace{0.4cm}
\begin{figure}[H]
	\centering
	\includegraphics[width=1\textwidth]{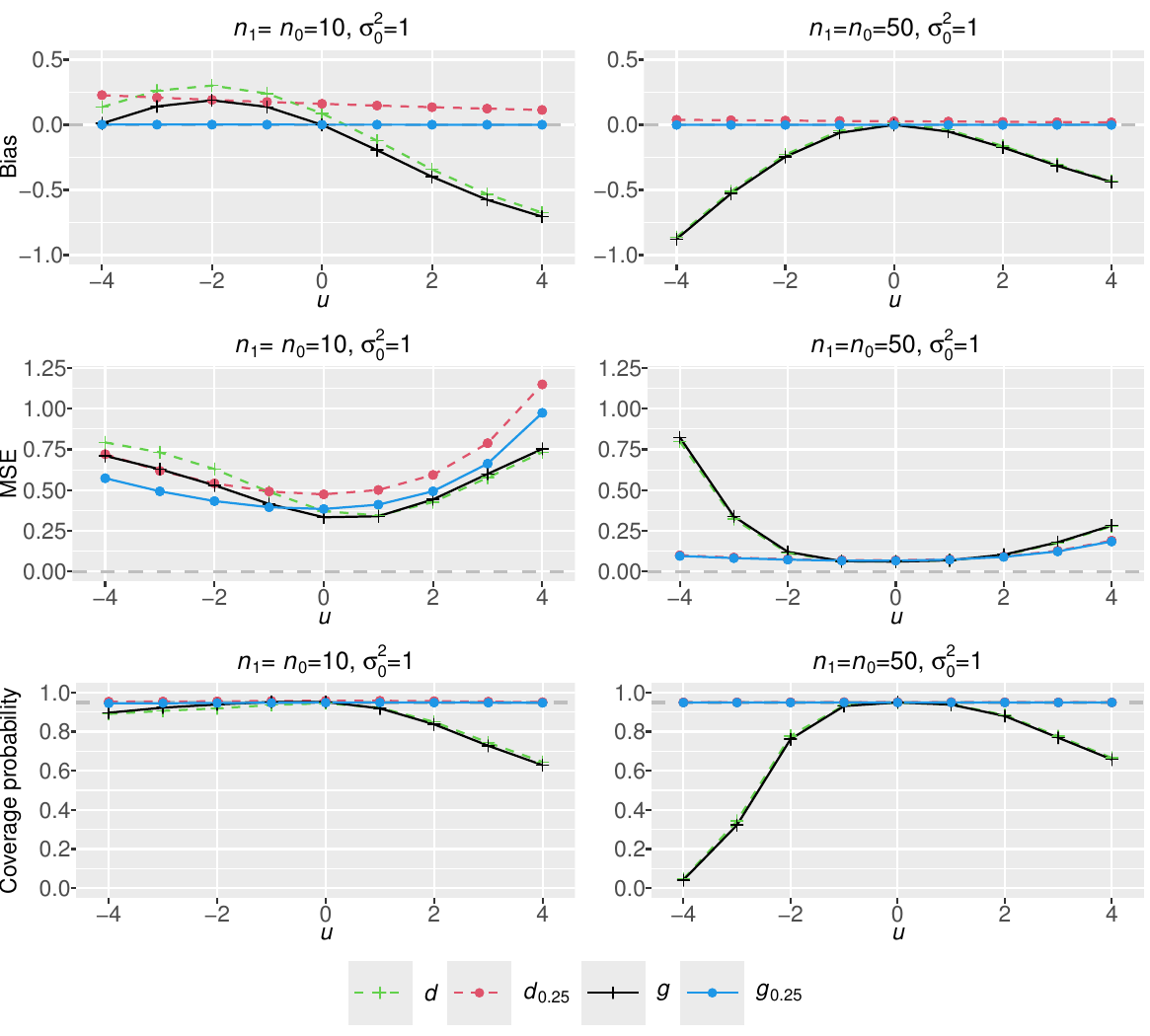}
	\setlength{\abovecaptionskip}{0pt}%
	\setlength{\belowcaptionskip}{0pt}%
	\caption{The bias, MSE and coverage probability for Cohen's $d$, our Cohen-type $d_w$ with $w=0.25$, Hedges' $g$, and our Hedges-type $g_w$ with $w=0.25$ are reported with $\sigma_0^2=1$ and $u$ ranging from -4 to 4, 
		where the left panels are for $n_1=n_0=10$,
		and the right panels are for $n_1=n_0=50$.}
	\label{fig:geo_pooled_2575}
\end{figure}

\vspace{0.4cm}
\begin{figure}[H]
	\centering
	\includegraphics[width=1\textwidth]{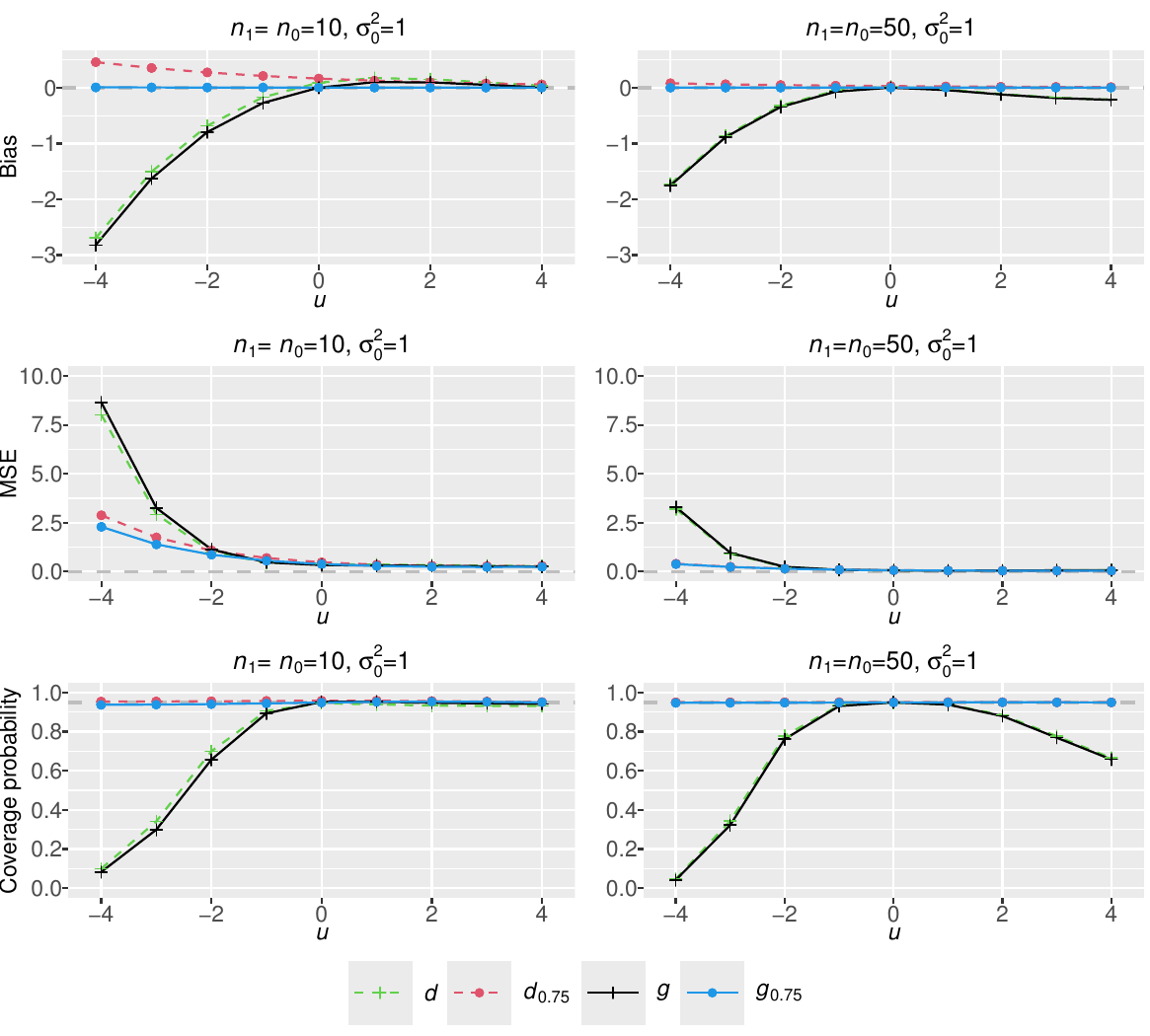}
	\setlength{\abovecaptionskip}{0pt}%
	\setlength{\belowcaptionskip}{0pt}%
	\caption{The bias, MSE and coverage probability for Cohen's $d$, our Cohen-type $d_w$ with $w=0.75$, Hedges' $g$, and our Hedges-type $g_w$ with $w=0.75$ are reported with $\sigma_0^2=1$ and $u$ ranging from -4 to 4, 
		where the left panels are for $n_1=n_0=10$,
		and the right panels are for $n_1=n_0=50$.}
	\label{fig:geo_pooled_2575_sample}
\end{figure}

\newpage
\begin{center}
	\textbf{\large Appendix F: The forest plot with the arithmetic SMD} 
\end{center}

\begin{figure}[H]
	\centering
	\includegraphics[width=1.2\textwidth,angle=90]{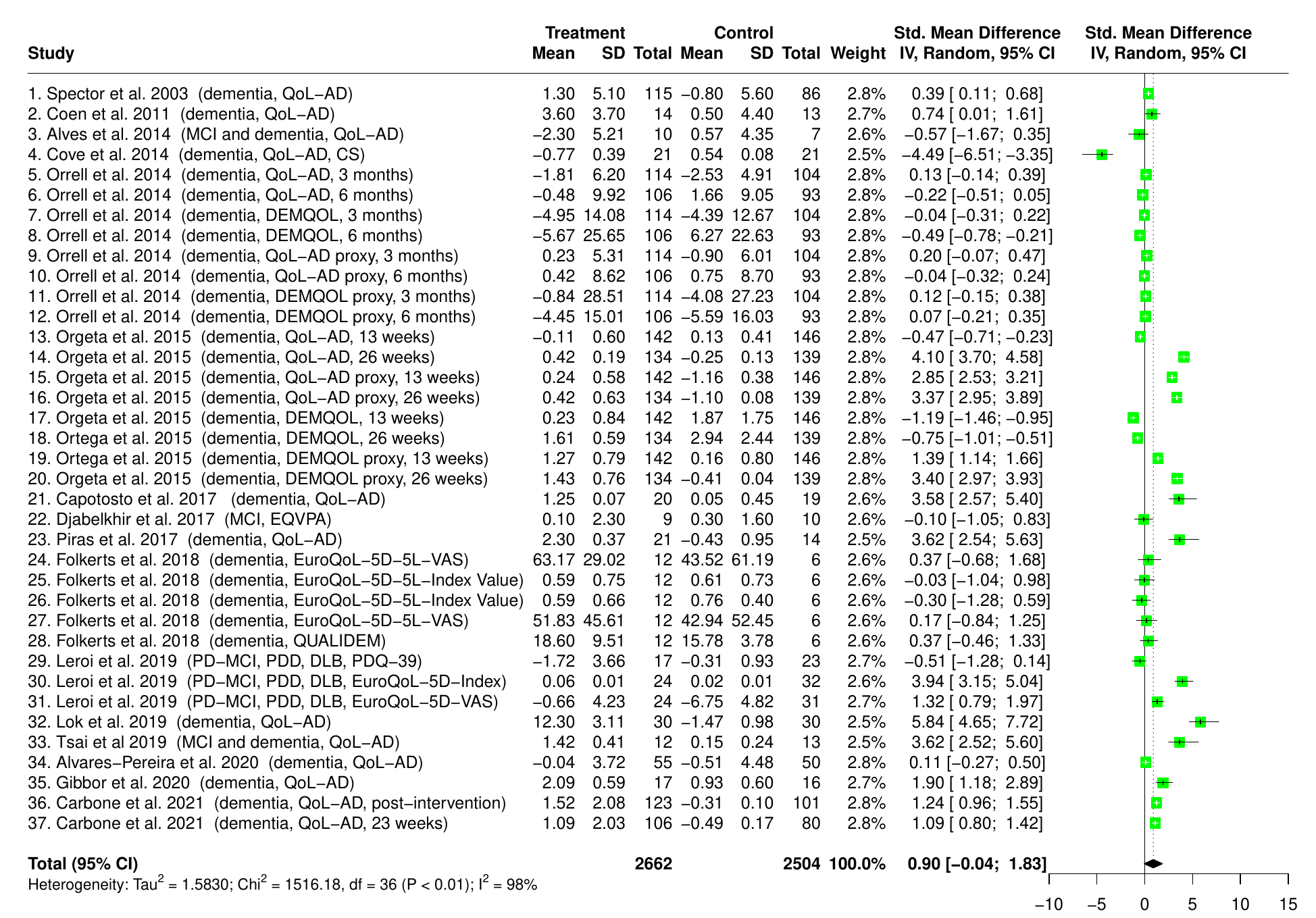}
	\setlength{\abovecaptionskip}{0pt}%
	\setlength{\belowcaptionskip}{0pt}%
	\caption{The meta-analytical results by the arithmetic SMD and its Hedges-type estimation.}
	\label{fig:arithmetic_results}
\end{figure}

\end{document}